\documentclass{article}

\usepackage{arxiv}

\usepackage[utf8]{inputenc} 
\usepackage[T1]{fontenc}    
\usepackage{hyperref}       
\usepackage{url}            
\usepackage{booktabs}       
\usepackage{amsfonts}       
\usepackage{nicefrac}       
\usepackage{microtype}      
\usepackage{lipsum}
\usepackage{graphicx}
\usepackage{amsmath}
\usepackage{bbm}
\graphicspath{ {./images/} }

\title{On Bayesian Generalized Additive Models}

\author{
  Antti Solonen
   \And
 Stratos Staboulis
}

\begin{document}
\maketitle
\begin{abstract}
Generalized additive models (GAMs) provide a way to blend parametric and non-parametric (function approximation) techniques together, making them flexible tools suitable for many modeling problems. For instance, GAMs can be used to introduce flexibility to standard linear regression models, to express "almost linear" behavior for a phenomenon. A need for GAMs often arises also in physical models, where the model given by theory is an approximation of reality, and one wishes to express the coefficients as functions instead of constants. In this paper, we discuss GAMs from the Bayesian perspective, focusing on linear additive models, where the final model can be formulated as a linear-Gaussian system. We discuss Gaussian Processes (GPs) and local basis function approaches for describing the unknown functions in GAMs, and techniques for specifying prior distributions for them, including spatially varying smoothness. GAMs with both univariate and multivariate functions are discussed. Hyperparameter estimation techniques are presented in order to alleviate the tuning problems related to GAM models. Implementations of all the examples discussed in the paper are made available.\footnote{\url{https://github.com/solbes/gam_paper_examples}}
\end{abstract}


\section{Introduction and notation}

Let us consider Generalized Additive Models of form 
\begin{equation}
    y_i = f_1(\mathbf{x}_{i,1}, \pmb{\theta}_1)+f_2(\mathbf{x}_{i,2}, \pmb{\theta}_2)+\cdots+f_M(\mathbf{x}_{i,M},\pmb{\theta}_M)+\varepsilon_i,
\end{equation}
where $f_i$ are functions, $y_i$ is observation, $\mathbf{x}_{i,j}$ are input observations needed in function $j$, $\pmb{\theta}_i$ are some unknown parameters for function $f_i$ and $\varepsilon_i$ are measurement errors.

One motivation for such models is to be able to include non-parametric effects to standard linear regression models; some of the terms above may be just $f_j(\mathbf{x}_{i,j}, \pmb{\theta}_j)_i=x_{i,j}\theta_j$, while some other terms may be treated as unknown functions, with the parameters being, for instance, the unknown values of the function at a chosen set of input points. Another motivation are "variable coefficient models", where one models the coefficients of a linear regression model as unknown functions:
\begin{equation}
\label{varcoef}
    y_i = f_1(\mathbf{z}_{i,1}, \pmb{\theta}_1)x_{i,1}+f_2(\mathbf{z}_{i,2}, \pmb{\theta}_2)x_{i,2}+\cdots+f_M(\mathbf{z}_{i,M},\pmb{\theta}_M)x_{i,1}+\varepsilon_i,
\end{equation}
where $\mathbf{z}_{i,j}$ are the required inputs for function $f_j$ for modeling observation $i$. With the above type of models, one can express the relationships between responses and inputs to be "almost linear", or the regression coefficients as "almost constant", and the model thus offers a framework for giving flexibility to standard linear regression models.

Note that the model described above differs a bit from the typical GAM formulation, where there is a nonlinear link function $g(y_i)$ attached to the observation. Here, we ignore this nonlinearity (by restricting to the case $g(y_i) = y_i$) and focus only on linear additive models, although many of the ideas discussed in the paper generalize to fitting such GAMs as well.

\subsection{General linear-Gaussian models}
\label{sec:lingaus}

In this paper we consider models where the functions $f_i$ are such that the final model with respect to the unknown parameters $\pmb{\theta}=[\pmb{\theta}_i,\pmb{\theta}_2, \cdots, \pmb{\theta}_M]^T$ is linear (but possibly non-linear with respect to model inputs). Furthermore, we want to restrict the unknown parameter values with Gaussian priors. That is, the whole model for estimating $\pmb{\theta}$ may always be written in the form
\begin{align}
\label{pos1}
\mathbf{y} &=  \mathbf{A}\pmb{\theta} + \pmb{\varepsilon} \\ 
\pmb{B\theta} &\sim \mathrm{N} \left( \pmb{\mu}_{\mathrm{pr}}, \pmb{\Gamma}_{\mathrm{pr}} \right),
\label{post}
\end{align}
where $\mathbf{y}=[y_1, \cdots, y_N]^T$ is the vector of all observations. Furthermore, assuming Gaussian measurement error, $\pmb{\varepsilon} \sim \mathrm{N} \left( \pmb{0}, \pmb{\Gamma}_{\mathrm{obs}} \right)$, we have a linear-Gaussian system with posterior $\pmb{\theta} | \mathbf{y} \sim \mathrm{N} \left( \pmb{\mu}_{\mathrm{pos}}, \pmb{\Gamma}_{\mathrm{pos}} \right)$, where
\begin{align} 
\pmb{\Gamma}_{\mathrm{pos}}^{-1} &= \mathbf{A}^{\mathrm{T}} \pmb{\Gamma}_{\mathrm{obs}}^{-1} \mathbf{A} + \mathbf{B}^{\mathrm{T}} \pmb{\Gamma}_{\mathrm{pr}}^{-1} \mathbf{B} \\ 
\pmb{\mu}_{\mathrm{pos}} &= \pmb{\Gamma}_{\mathrm{pos}} \left( \mathbf{A}^{\mathrm{T}} \pmb{\Gamma}_{\mathrm{obs}}^{-1} \mathbf{y} + \mathbf{B}^{\mathrm{T}} \pmb{\Gamma}_{\mathrm{pr}}^{-1} \pmb{\mu}_{\mathrm{pr}} \right).
\end{align}
Note that $\mathbf{A}$ and $\mathbf{B}$ are often sparse matrices, and noise covariances are often taken to be diagonal, so the above system can be efficiently solved with sparse linear algebra methods even with high dimensional data and parameters. Moreover, the posterior precision matrix $\pmb{\Gamma}_{\mathrm{pos}}^{-1}$ is often also sparse and can be used to, e.g., efficiently sample from the Gaussian posterior.

The mapping from various formulations for $f_i$ to the above system is discussed further on in the paper. For instance, classical linear regression is obtained by setting $f_j(\mathbf{x}_{i,j}, \pmb{\theta}_j)_i=x_{i,j}\theta_j$, and then we have simply
\begin{equation}
    \pmb{A}=
    \begin{bmatrix}
        \mathbf{x}_1 & \mathbf{x}_2 & \cdots & \mathbf{x}_M
    \end{bmatrix},
\end{equation}
where $\mathbf{x}_j=[x_{1,i},\ x_{2,j},\ \cdots,\ x_{N, j}]^T$, which is the typical design matrix in linear regression.

The linear transformation $\pmb{B}$ for the prior system is included for the cases, where we don't want to penalize the unknowns directly, but, e.g., the differences between consecutive unknown function values in a spatial grid. Note that $\pmb{B}$ does not need to be full rank; the prior can be improper as long as the above posterior is proper. This is the case, for instance, if we only set priors for differences between consecutive function values, but don't penalize the function values directly.



\subsection{Basis function representation}
\label{basis}

In GAMs, we often wish to estimate the functions $f_i$ using non-parametric techniques. For that purpose, in most cases, we are able to write the functions using a linear combination of chosen basis functions:
\begin{equation}
    (f_i)_j = \sum_{k=1}^{K_i} \alpha_{i,k} p_{i,k}(\mathbf{x}_{i,j}),
\end{equation}
where the unknown parameters related to the function are the weights, $\pmb{\theta}_i=[\alpha_1, ..., \alpha_K]$. For instance, the basis functions could be a set of spline functions defined at a selected set of knot points.

With the above basis function representation of the unknown functions, we arrive at a "stacked form" linear system, where each $f_i$ introduces $K_i$ columns to the system matrix $\mathbf{A}$:

\begin{equation}
    \mathbf{A} = 
    \begin{bmatrix}
        \overbrace{p_{1,1}(\mathbf{x}_{1,1})\ \ \cdots\ \ p_{1,K_1}(\mathbf{x}_{1,1})}^{f_1} & \overbrace{p_{2,1}(\mathbf{x}_{2,1})\ \ \cdots\ \ p_{2,K_2}(\mathbf{x}_{2,1})}^{f_2} & \cdots & \overbrace{p_{M,1}(\mathbf{x}_{M,1})\ \ \cdots\ \ p_{M,K_M}(\mathbf{x}_{M,1})}^{f_M} \\
        
        p_{1,1}(\mathbf{x}_{1,2})\ \ \cdots\ \ p_{1,K_1}(\mathbf{x}_{1,2}) & p_{2,1}(\mathbf{x}_{2,2})\ \ \cdots\ \ p_{2,K_2}(\mathbf{x}_{2,2}) & \cdots & p_{M,1}(\mathbf{x}_{M,2})\ \ \cdots\ \ p_{M,K_M}(\mathbf{x}_{M,2}) \\
        
        \vdots & \vdots & \vdots & \vdots \\
        
        p_{1,1}(\mathbf{x}_{1,N})\ \ \cdots\ \ p_{1,K_1}(\mathbf{x}_{1,N}) & p_{2,1}(\mathbf{x}_{2,N})\ \ \cdots\ \ p_{2,K_2}(\mathbf{x}_{2,N}) & \cdots & p_{M,1}(\mathbf{x}_{M,N})\ \ \cdots\ \ p_{M,K_M}(\mathbf{x}_{M,N}) \\
        
        & & & 
        
    \end{bmatrix}
\end{equation}

Sometimes it makes sense to add a translation to the basis function representation, in which case we would have
\begin{equation}
    (f_i)_j = m_{i}(\mathbf{x}_{i,j}) + \sum_{k=1}^{K_i} \alpha_{i,k} p_{i,k}(\mathbf{x}_{i,j}),
\end{equation}
where $m_{i}$ is the translation of function $i$; each $m_{i}$ is a known function of the input data. In this case, the $\mathbf{A}$ matrix stays the same as in the above non-translated version, but the translations need to be subtracted from the observation data before model fitting to get the system in the common form given in (\ref{pos1})-(\ref{post}).

Note also that the above extends trivially to the "variable coefficient models" given in equation (\ref{varcoef}): one simply multiplies the basis vectors with the appropriate input data.

\section{Literature review}

The book \cite{wood17} is perhaps the most substantial text on the subject, together with the recent paper \cite{wood20} that discusses some recent developments in GAMs. They cover all aspects of GAMs and introduce also \texttt{R} code for computations. The text is written mainly from the frequentist perspective, but also briefly discusses the Bayesian viewpoint, and, for example, draws a connection between penalized smoothing splines and Gaussian smoothness priors.

The GAMs in \cite{wood17, wood20} focus on local basis functions described via smoothing splines. In this paper we discuss both local and global basis functions, and describe some novel features, such as spatially varying smoothness, dimension reduction and arbitrary orders for the smoothness priors (in the previous work the order is always two). In addition, we present Bayesian methods for estimating the GAM hyperparameters.

The RSS paper \cite{rue09} discusses fitting GAM models with INLA (integrated nested Laplace approximations). The paper discusses hyperparameter estimation and also various non-Gaussian likelihoods, and how to efficiently approximate the posterior distribution. Gaussian Markov random fields (GRMFs) are chosen as priors. The cited paper focuses mainly on the inference methodology in non-Gaussian problems, whereas in this paper we restrict to the linear-Gaussian case and focus the discussion on how to express GAMs using various local and global basis function constructions.

Multivariate adaptive regression splines (MARS), see \cite{hastie09}, is another approach for building GAMs. In MARS, the basis functions are either constants or "hinge functions" and the approach can be extended to multiple dimensions via products of hinge functions. The method works by first adding terms in a greedy way and then pruning the model by removing terms with cross-validation. In contrast to the present paper, the viewpoint is not Bayesian. In addition, we argue that in the MARS approach it is rather difficult to control the smoothness and other properties of the model (e.g. periodicity), which is one of the main topics of this paper.

\section{Global basis functions via Gaussian Processes}
\label{sec:gp}

One popular non-parametric technique for function approximation is Gaussian Processes. A Gaussian process (GP) is an infinite dimensional (function space) concept, and can loosely be defined as such a stochastic process, where the distribution of any finite collection of realizations of the function values is a multivariate normal distribution. We discuss here GPs in the context of GAMs; for a thorough treatment of GPs in general, refer to, e.g., \cite{rasmussen06}.

A GP is characterised by its mean function $m(x)$ and covariance function (or kernel) $k(x,x')$. For practical computations, one can discretize the GP onto a chosen grid of input values, and then the GP turns into a multivariate Normal distribution.

Here, we present the functions in our GAM model as GPs:
\begin{equation}
    f_i(\mathbf{x}, \pmb{\theta}) \sim \mathcal{GP} \left( m_i(\mathbf{x}), k_i(\mathbf{x}, \mathbf{x}') \right).
\end{equation}
where the definition of the model parameters $\pmb{\theta}$ is specified further below.
In practice, we discretize the function to a fixed grid of input points, and thus $\mathbf{f}_i \sim N(\pmb{\mu}_{f_i}, \pmb{\Sigma}_{f_i})$, where the mean vector and covariance matrix are calculated via the chosen mean and covariance functions.

To use our basis function representation for GAMs here, let us take the eigenvalue decomposition for the covariance matrix:
\begin{equation}
     \pmb{\Sigma}_f = \sum_{i=1}^n \lambda_i \mathbf{q}_i\mathbf{q}_i^{\mathrm{T}} = \mathbf{P}\mathbf{P}^{\mathrm{T}},
\end{equation}
where column $i$ of the matrix $\mathbf{P}$ is the $i$:th eigenvector scale by the corresponding eigenvalue: $\mathbf{p}_i=\sqrt{\lambda_i}\mathbf{q}_i$. It's easy to verify that $\mathbf{f}_i = \pmb{\mu}_{f_i} + \mathbf{P}\pmb{\theta}$ with $\pmb{\theta} \sim N(\mathbf{0},\mathbf{I})$. Now the basis functions are the columns of $\mathbf{P}$, the model parameters $\pmb{\theta}$ are the weights for the basis vectors (with a convenient i.i.d. Gaussian prior), the translation is given by $\pmb{\mu}_{f_i}$, and we can then use the basis vector representation discussed in Section \ref{basis} for computation. Note, however, that the functions are here defined on a grid of input points, but in practice we need to evaluate the functions at the observation locations. This can be done by simply interpolating to these locations.

In infinite-dimensional terms, we are aiming to use the eigenfunctions as the basis vectors for GAMs here. The discretization-interpolation trick described above is one way to approximate the eigenfunctions in practice. There are, however, other methods, such as the Nyström method, see \cite{marzouk, press92}. In addition, for some kernels, we might have the eigenfunctions available analytically \cite{rasmussen06}.

With this approach, the number of obtained basis vectors (and the number of additional unknowns) equals the number of grid points chosen for the GP representation. However, there is a straightforward way of reducing the dimension of the estimation problem by dropping out the basis vectors that do not contribute much to the covariance. We can choose only the first $k < n$ eigenvalues and approximate $\pmb{\Sigma}_f \approx \mathbf{P}_k\mathbf{P}_k^{\mathrm{T}}$, where $\mathbf{P}_k$ contains the basis vectors corresponding to the $k$ largest eigenvalues, so that the truncated covariance contains, e.g., 99.99\% of the original covariance\footnote{One can select the eigenvectors $k$ for which $(\sum_{i=1}^k \lambda_i)/(\sum_{i=1}^n \lambda_i) < 0.9999$, for instance.}. Now we can re-parameterize accordingly: $\mathbf{f} = \pmb{\mu}_{f} + \mathbf{P}_k\pmb{\theta}$ and estimate only $k$ weights instead of the original $n$. This is a well-known dimension reduction approach (the {\em "truncated SVD"} method) in inverse problems among which one typically encounters high-dimensional estimation problems; for example when the unknown is, e.g., a spatial function, see \cite{marzouk}. The success of the dimensionality reduction technique depends on how smoothing the selected prior GP is; the smoother the unknown function is assumed, the lower dimension is needed in the estimation. This approach gets particularly useful when the number of (independent) input variables is greater than 1, as discussed in the next section.

\subsection{GP bases in multiple dimensions}

Since we express the GP in a chosen grid, the total number of unknowns grows rapidly as a function of the dimensionality of the input space. For instance, using 50 grid points in each direction for a trivariate GP would require $50^3=125000$ points. Even storing such high dimensional covariance matrices is impossible. Also, leveraging sparse matrix algebra is usually hard, since most of the typically used covariance kernels result in dense covariance matrices.

Restricting to \textit{separable} covariance functions, coupled with the dimensionality reduction approach described in the previous section, provides a way forward. A covariance function is said to be separable, if it can be written as a product of lower-dimensional covariance functions. For example, a two dimensional separable covariance function would be $k(\mathbf{x},\mathbf{x}')=k(x_1,x_1' )k(x_2,x_2' )$. Let us now discretize the two variables so that we get $n$ grid points for $x_1$ and $m$ grid points for $x_2$, and calculate the one-dimensional covariance matrices $\mathbf{K}_1 \in \mathbb{R}^{n \times n}$ and $\mathbf{K}_2 \in \mathbb{R}^{m \times m}$. Now the full covariance matrix in the joint space can be calculated as the Kronecker product of the individual covariances (assuming a suitable ordering of the variables):
\begin{equation}
    \mathbf{K}=\mathbf{K}_1 \otimes \mathbf{K}_2 \in \mathbb{R}^{mn \times mn} .
\end{equation}

How about the basis vectors in the joint space? We can use the properties of Kronecker products, see, e.g., \cite{laub05}, to calculate the eigen-decomposition of the full matrix using the decompositions of the lower dimensional matrices, without having to form $\mathbf{K}$ explicitly. In the two-dimensional example, if the one-dimensional covariances have decompositions $\mathbf{K}_1 = \mathbf{Q}_1 \pmb{\Lambda}_1 \mathbf{Q}^{\mathrm{T}}$ and $\mathbf{K}_2 = \mathbf{Q}_2 \pmb{\Lambda}_2 \mathbf{Q}_2^{\mathrm{T}}$, the two-dimensional covariance has decomposition
\begin{equation}
    \mathbf{K}=\mathbf{K}_1 \otimes \mathbf{K}_2 = (\mathbf{Q}_1 \otimes \mathbf{Q}_2) (\mathbf{\Lambda}_1 \otimes \mathbf{\Lambda}_2) (\mathbf{Q}_1 \otimes \mathbf{Q}_2)^{\mathrm{T}}.
\end{equation}
That is, the eigenvectors are Kronecker-products of the individual eigenvectors, and eigenvalues are products of the eigenvalues (after a proper ordering of the vectors and values). For dimensions higher than 2 the decomposition works analogously.

In this way we can apply the dimensionality reduction technique, for each individual input variable, to take into account only eigenvectors that significantly contribute to the prior covariance while discarding the insignificant ones. This procedure can drastically reduce the overall dimensionality of the joint space of estimated unknowns.

\subsection{Some useful kernels and examples}

Perhaps the most useful kernel in the GAM context is the squared exponential kernel
\begin{equation}
    k(\mathbf{x}, \mathbf{x}') = \sigma^2 \exp \left( -\frac{d(\mathbf{x}, \mathbf{x}')^2}{2L^2} \right),
\end{equation}
where $d(\mathbf{x}, \mathbf{x}')$ is the Euclidean distance, $\sigma^2$ is variance and $L$ is correlation length. The variance roughly describes how far from the mean the function can fluctuate, and the correlation length determines the smoothness of the function, in other words, how similar two function values that are a certain distance apart are. The kernel yields very smooth realizations, and the eigenvalues of the kernel decay quickly, which makes it suitable for the dimensionality reduction method discussed above. 

Sometimes we need to describe periodic functions. For example, when dealing with functions of angles or time periods. A periodic version of the squared exponential kernel can be written as
\begin{equation}
    k_{\mathrm{periodic}}(\mathbf{x}, \mathbf{x}') = \sigma^2 \exp \left( \frac{-2\sin^2 ( \pi d(\mathbf{x}, \mathbf{x}') / \phi)}{L^2} \right),
\end{equation}
where $\phi$ is the period. Moreover, we can often assume symmetry in the underlying function; $f(x)=f(-x)$. Such symmetric kernels can be constructed from existing kernels $k(x,x')$ via 
\begin{equation}
    k_{\mathrm{sym}}(x, x') = k(x,x') + k(-x,x').
\end{equation}

\begin{figure}
    \centering
    \includegraphics[width=\textwidth]{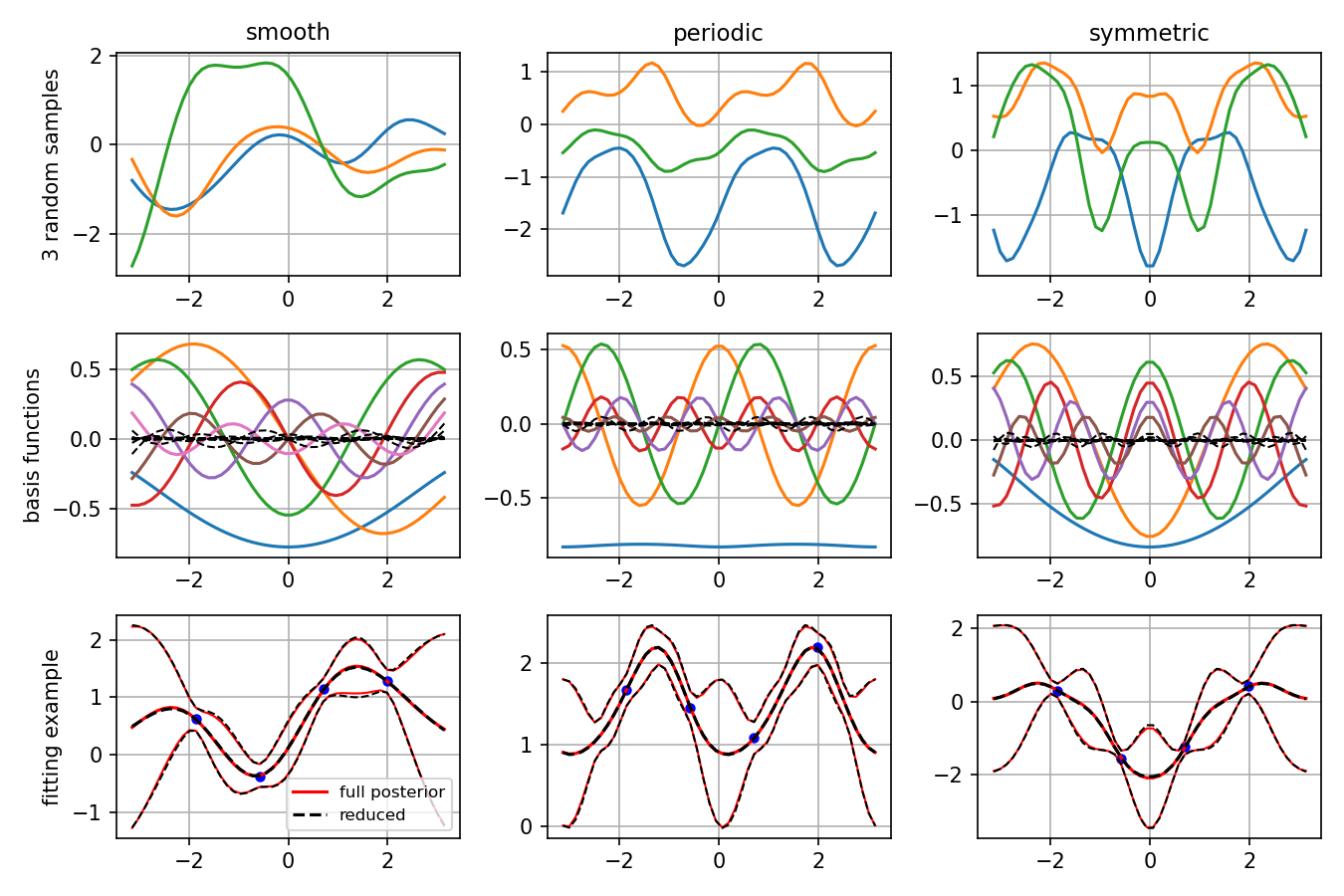}
    \caption{Random samples (top row), basis vectors (middle row) and simple fitting examples (bottom row) for the squared exponential kernel (first column) and its periodic (middle column) and symmetric (last column) versions. Black dashed lines in the second row plots illustrate the basis functions that are dropped out from the estimation. The last row compares the mean and +- two standard deviations calculated with the full and reduced bases.}
    \label{fig:gp_basis_demo}
\end{figure}

An example of the discussed kernels and their application in the reduced rank basis function context is given in Fig. \ref{fig:gp_basis_demo}. One can see that using the dimension reduction approach with a strongly smoothing kernel enables using only a small number of basis vectors to model the functions.

\section{Local basis functions}

By far the most common formulation of GAMs involve splines for representing the unknown functions in a non-parametric way, see, e.g., \cite{wood17} for a thorough treatment of splines in the context of GAMs. In spline-based GAMs, the unknowns are the coefficients of the spline basis functions. 

Smoothness in the functions can be obtained with two mechanisms: 1) by the number of chosen knot locations (lower number of knots gives smoother functions) and 2) explicitly penalizing the smoothness of a function by, e.g., adding a penalization term for the second derivative: $\lambda \int f''(x)dx$, where $\lambda$ is a tuning parameter (hyperparameter). The derivative can be analytically calculated for splines, and results in a quadratic penalty for the unknown spline coefficients. In Bayesian terms, this is equivalent to a Gaussian prior put on the spline coefficients.

The estimation problem arising from spline-based GAMs can be high-dimensional. For instance, typically it makes sense to select a "little bit too fine" grid of knot locations, and then use the smoothness penalty to select the appropriate smoothness level for the function. However, the difficulty is alleviated by the fact that the spline basis functions are non-zero only close to the knot location in question and zero elsewhere, which makes it possible to use sparse linear algebra in the computations.

\subsection{First order splines and difference priors in 1D}

Let us here discuss the first order B-splines as building blocks for GAMs. This is equivalent to simply treating the function values directly as the unknowns, and performing linear interpolation to the regions in between the seleted input points. In our estimation framework, this means that the matrix $\mathbf{A}$ is the sparse linear interpolation matrix that maps given function values at the selected input grid to the observation locations. An example of $\mathbf{A}$ is given below:
\begin{equation}
    \mathbf{A} = \begin{bmatrix}
0 & \cdots & w_{1} & 1-w_{1} & \cdots & \cdots & \cdots & \cdots & 0 \\
0 & \cdots & \cdots & \cdots & \cdots & w_{2} & 1-w_{2} & \cdots & 0 \\
\vdots & \vdots & \vdots & \vdots & \vdots & \vdots & \vdots & \vdots & \vdots
\end{bmatrix},
\end{equation}
where $w_{i}$ is the distance of the input observation to the first input grid value that it exceeds. The matrix elements, using the input grid points $\mathbf{x}_g = [x_{g,1}, x_{g,2}, ..., x_{g,N}]$, are thus given as
\begin{equation}
    A_{ij} = \begin{cases}
    x_i-x_{g,j} & \mathrm{if}\ x_i \in [x_{g,j}, x_{g,j+1}]\\
    1-(x_i-x_{g,j}) & \mathrm{if}\ x_i \in [x_{g,j-1}, x_{g,j}]\\
    0 & \mathrm{otherwise}
    \end{cases}
\end{equation}

Instead of just focusing on second order smoothness penalties with single tuning parameters, let us take a bit wider perspective here. We could, of course, directly restrict the function values with Gaussian priors. Often, however, it is more useful to penalize the changes in the function values (derivatives). Approximating the derivatives in the chosen grid leads to sparse difference matrices. The order of the difference matrix does not necessarily have to be two, as typically done with spline-based GAMs. Below are examples of the first three orders of difference matrices for a one-dimensional input grid of length 5:
\begin{equation}
    \mathbf{D}_1 = 
    \begin{bmatrix}
        -1 &  1 & 0 & 0 & 0 \\
        0 &  -1 & 1 & 0 & 0 \\
        0 &  0 & -1 & 1 & 0 \\
        0 &  0 & 0 & -1 & 1
    \end{bmatrix}
    \ \ 
    \mathbf{D}_2 = 
    \begin{bmatrix}
        -1 &  2 & -1 & 0 & 0 \\
        0 &  -1 & 2 & -1 & 0 \\
        0 &  0 & -1 & 2 & -1
    \end{bmatrix}
    \ \ 
    \mathbf{D}_3 =
    \begin{bmatrix}
        -1 &  3 & -3 & 1 & 0 \\
        0 &  -1 & 3 & -3 & 1
    \end{bmatrix}
    .
\end{equation}
Note that we ignore the "boundary conditions" for the difference operators here, and only penalize the differences, not the function values. When used as $\mathbf{B}$ matrices in our fitting formulation, these lead to improper priors. However, this is fine as long as one has enough observation data to get a proper posterior.

\begin{figure}
    \centering
    \includegraphics[width=\textwidth]{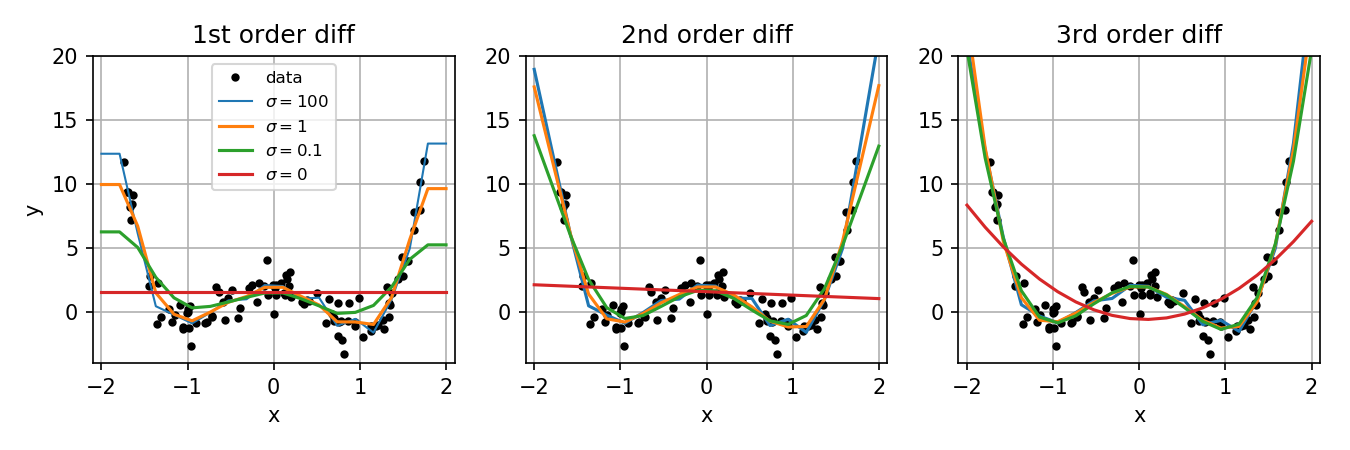}
    \caption{Fitting data generated around the true function $f(x)=3x^4-6x^2+2$ using three different orders for the difference priors and four different prior variances.}
    \label{fig:diff_demo}
\end{figure}

Different orders for the difference priors behave a bit differently, see Fig. \ref{fig:diff_demo} for an example of using first, second, and third order differences for a simple function approximation task. One can observe that while all of the choices give similar fits for the data range, the extrapolation behavior is different. The first order prior tries to keep the behavior constant, second order keeps it linear, and third order extrapolates quadratically. It depends on the application which of these is the most suitable.

Note that sometimes the best solution for estimating a function within a GAM is to have a mixture of various orders of difference priors; we might, for instance, know that a one-dimensional function is expected to extrapolate linearly to the other direction, but saturate to some constant value at the other end.

\subsection{Functions in higher dimensions}

Spline bases can be extended to higher dimensions via "tensor product bases", see \cite{wood17} for discussion. Let us here focus on the first order splines and difference priors discussed in the previous section. The linear interpolation model can be extended to multiple dimensions in a straightforward manner; one needs to decide a suitable ordering, vectorize the high-dimensional function values according to the chosen ordering, and construct the interpolation matrix accordingly (calculate differences between the correct vector indices). For instance, the first row of a two-dimensional interpolation matrix using column ordering would have 4 non-zero values and look like
\begin{equation}
    \mathbf{A}_{1\cdot} = \left[
0\ \ \cdots\ \ 0\ \ w_{11}\ \ w_{12}\ \ 0\ \ \cdots\ \ 0\ \ w_{13}\ \ w_{14}\ \ 0\ \ \cdots\ \ 0 
\right],
\end{equation}
where $w_{1i}$'s represent the distances to the corners of the "square pixel" where the first observation falls in, and $\sum_i w_{1i}=1$. Each row in an N-dimensional interpolation matrix would have $2^N$ non-zero elements.

The difference priors of a given order are also extendable to higher dimensions in an analogous way. For constructing the priors, it might make sense to define the priors separately to the direction of each dimension (one might even want to specify a different order for the differences for different dimensions). As an example, the second order difference matrices in the above two-dimensional example for both directions are given below:
\begin{equation}
    \mathbf{D}_1 = 
    \begin{bmatrix}
        -1 &  0 &  0 & 2 &  0 &  0 &  -1 &  0 &  0 \\
        0 &  -1 &  0 & 0 &  2 &  0 &  0 &  -1 &  0 \\
        0 &  0 &  -1 & 0 &  0 &  2 &  0 &  0 &  -1
    \end{bmatrix}
    \ \ 
    \mathbf{D}_2 = 
    \begin{bmatrix}
         -1 &  2 &  -1 & 0 &  0 &  0 &  0 &  0 &  0 \\
        0 &  0 &  0 & -1 &  2 & -1 &  0 &  0 &  0 \\
        0 &  0 &  0 & 0 &  0 &  0 &  -1 &  2 &  -1
    \end{bmatrix}.
\end{equation}
The final prior matrix $\mathbf{B}$ in this case could be the combination of the priors in both dimensions. Code for generating the N-dimensional sparse interpolation matrix and the difference prior matrices is made available in the attached package\footnote{\url{https://github.com/solbes/blinpy}}.

\begin{figure}
    \centering
    \includegraphics[width=0.7\textwidth]{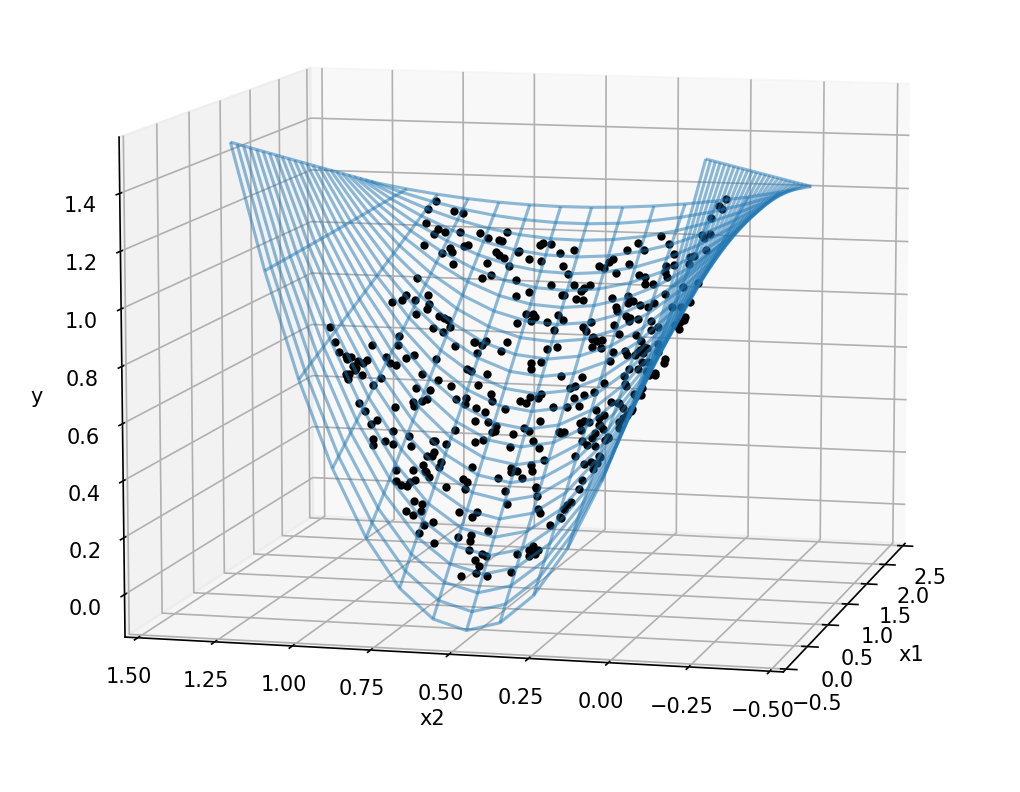}
    \caption{Fitting data generated around the true function $f(x_1, x_2)=0.5x_1+4(x_2-0.5)^2/(1+2x_1)$ using second order difference prior in the first dimension and third order difference prior in the second dimension.}
    \label{fig:diff_demo_3d}
\end{figure}

Let us consider an example of fitting data randomly generated around the true function $f(x_1, x_2)=0.5x_1+4(x_2-0.5)^2/(1+2x_1)$. Here, the function is quadratic in the second dimension, with curvature varying as a function of $x_1$, and behaves rather linearly in the first dimension. Let us fit here a function with second order difference prior in the first dimension and third order prior in the second dimension. The result is illustrated in Fig. \ref{fig:diff_demo_3d}; one can see that the fit extrapolates nicely along both dimensions.

The approach of using difference priors in approximating unknown functions is a widely studied topic in the field of inverse problems. For instance, \cite{haario04} discusses difference priors in atmospheric remote sensing applications, and shows how to make the priors discretization independent. See also \cite{bardsley13, rue05} and the references therein for discussion about Gaussian Markov Random Fields and their connection to difference priors in the inverse problems context.

\subsection{Spatially varying smoothness}

\begin{figure}
    \centering
    \includegraphics[width=\textwidth]{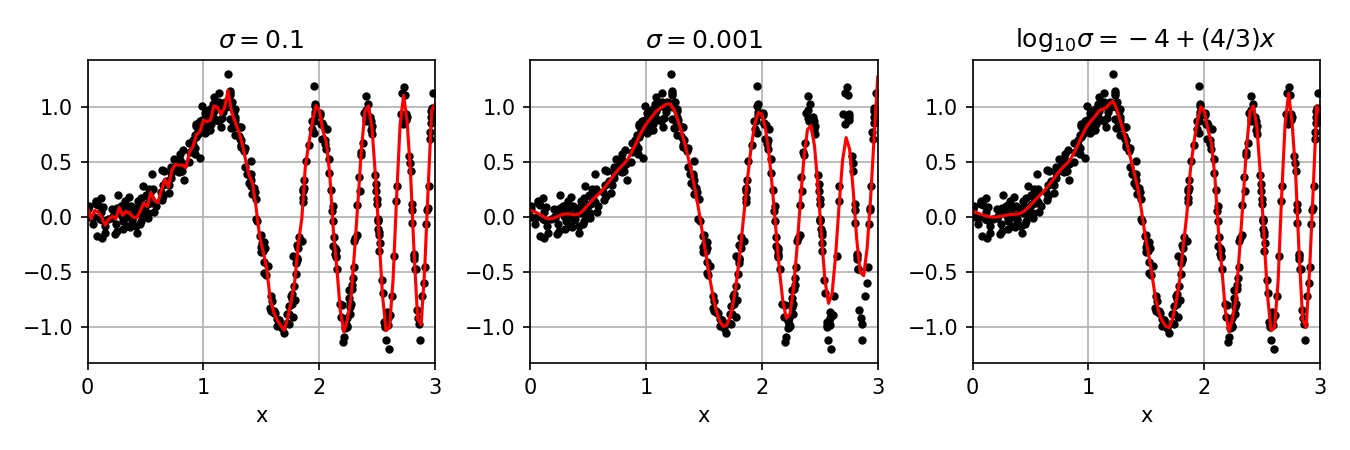}
    \caption{Fitting data generated around the true function $f(x)=\sin(x^3)$ using three different three different ways for defining the prior variance. For the first two plots, the prior variance is fix to a constant value, but for the last figure, the prior variance is increased as a function of $x$.}
    \label{fig:spatial_demo}
\end{figure}

With the local basis approach, there is another level of flexibility that is typically not utilized in GAM approaches. By tuning the prior variances of the difference priors, one can easily achieve spatially varying smoothness in the functions. An example of this is considered in Fig. \ref{fig:spatial_demo}, where the goal is to fit a function that clearly has non-constant smoothness. Specifying a prior variance that works for one part of the function might work very poorly at some other parts of the function. Defining different variances for different parts of the function solves the problem. Note that here the behavior of the variance is chosen manually. Automatically estimating a spatially varying smoothness parameter is tricky, see Section \ref{sec:hyperpar} for more discussion.

\begin{figure}
    \centering
    \includegraphics[width=0.7\textwidth]{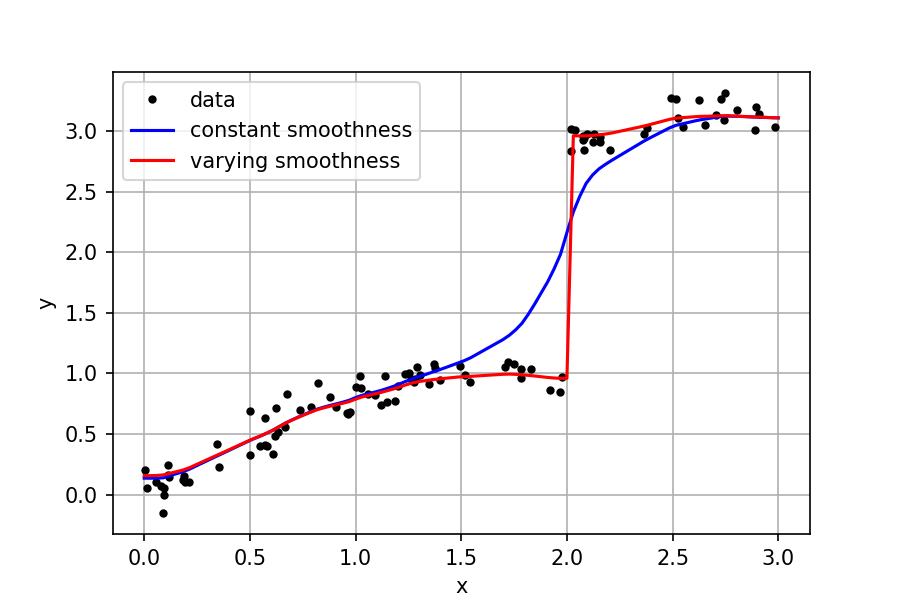}
    \caption{Fitting random data generated around the true function $f(x)=\sin(x)+\mathbbm{1}(x>2)$ by 1) keeping the prior variance constant, 2) increasing the prior variance around $x=2$.}
    \label{fig:sharp_demo}
\end{figure}

Another typical application for spatially varying smoothness is modeling sudden jumps in the functions. Such a behavior can be obtained by setting the prior variances high in the vicinity of the discontinuity, see \cite{Bardsley_2010} for discussion in the context of inverse problems, where boundary preserving priors arising, e.g., in medical imaging applications, is a topic of active research. Many of those ideas generalize to GAMs as well. An example is given in Fig. \ref{fig:sharp_demo}, where the data contains a large jump, and by controlling the spatially varying smoothness one can obtain a good fit to the data.

\subsection{Resolving the identifiability issue}

With the local basis function approach, where only the derivatives of the function are penalized, there is an obvious identifiability issue when using multiple additive components in the same model. For instance, for two additive components, the functions are only identifiable up to a constant; one can add a constant value to the first function and substract the values from the second, and the model predictions are exactly the same, as are the penalty terms (function derivatives). Thus, more prior information is required in the model to make it identifiable. Another option would be to use the Moore-Penrose pseudo-inverse in the model fitting to obtain the minimum norm solution to the linear problem.


A straightforward way to resolve the issue in the Bayesian setting is to set a weakly informative Gaussian prior (mean and variance) separately for each function, $\theta_i \sim N(\overline{\theta}_i, \tau_i)$. The means and variances for the functions could be obtained manually, or then one could first fit a model where the functions are assumed constants, and allow some deviation around these values (e.g., standard deviation could be chosen to be a percentage of the estimated constant parameter values).

If the model is not a variable coefficient model, where the functions are multiplied with additional inputs, one can also include a common "mean coefficient" $\overline{\theta}$ and require that all the estimated functions are around this mean coefficient by setting $\theta_i \sim N(\overline{\theta}, \tau)$. The mean coefficient can be estimated along with the other parameters, or taken to be the mean of the data, for instance. The standard deviation could be taken to be a multiple of the standard deviation of the data, for instance. Note that choosing $\overline{\theta}=0$ gives the minimum norm solution as the MAP estimate of the resulting problem.

The approach used in \cite{wood17} for resolving the issue is to add constraints to the estimation problem so that the sum of a smooth component should sum up to zero, and then add an intercept term to the model that picks up the correct level. The benefit of this is that the confidence bands for the smooth functions stay "narrow". However, this does not directly work with variable coefficient models; one would need to add several constant parameters multiplied with the appropriate inputs.

Note that the GP parameterisation discussed in section \ref{sec:gp} does not suffer from this identifiability issue, since the GP prior penalizes the function values directly and each function tends towards their own chosen mean values.

\subsection{Periodic and symmetric priors}

In the GP prior case, requiring periodicity and symmetry was obtained in a straightforward manner by choosing a suitable kernel with the desired periodic/symmetric properties. With the local basis functions and difference prior approach, one needs to include additional rows in the difference prior matrices to obtain such behavior.

Periodicity in the function values can be obtained simply by introducing one extra row in the prior system that calculates the difference between the first and the last function values. However, in addition to matching the function values at the boundaries, one would often like to match some derivatives too. Requiring that the function values and first derivatives match at the boundary in a one-dimensional case would lead to the following rows to be added to the prior system:
\begin{equation}
    D_{periodic} = \begin{bmatrix}
         -1 & 0 & 0 & \cdots & 0 & 0 & 1 \\
         1 & -1 & 0 & \cdots & 0 & -1 & 1
    \end{bmatrix}.
\end{equation}
The "strength" of the periodicity assumption can be controlled with the associated prior variances. Custom periods for the periodicity priors can be obtained in an analogous manner by calculating the differences between the correct function values. This requires a bit more manual work than with the GP approach, where the period is explicitly a parameter in the covariance kernel. Then again, difference priors offer more flexibility as one doesn't need to specify any restrictive covariance kernel explicitly.

In addition to periodicity, symmetry is a common assumption that can be made of the underlying function. Symmetry can be enforced by penalizing the correct differences calculated on both sides of the axis of symmetry. For instance, for a 6-dimensional input grid, symmetry about the middle of the domain can be enforced by setting the prior system matrix to
\begin{equation}
    \mathbf{B} = \begin{bmatrix}
          0 &  0 & -1 &  1 &  0 &  0 \\
          0 &  -1 & 0 &  0 &  1 &  0 \\
          -1 &  0 & 0 &  0 &  0 &  1 
    \end{bmatrix}.
\end{equation}

An example of fitting symmetric and periodic functions is given in Fig \ref{fig:symmetric_demo} below. In the first figure, we fit the model using only smoothness prior. In the second figure we add a periodic prior, and one can see that the values and derivatives at the end of the domain match. In the last figure we add a prior that the function needs to be symmetric, which further improves the fit. Note that in the last figure the symmetry is not about the center of the domain.

\begin{figure}
    \centering
    \includegraphics[width=\textwidth]{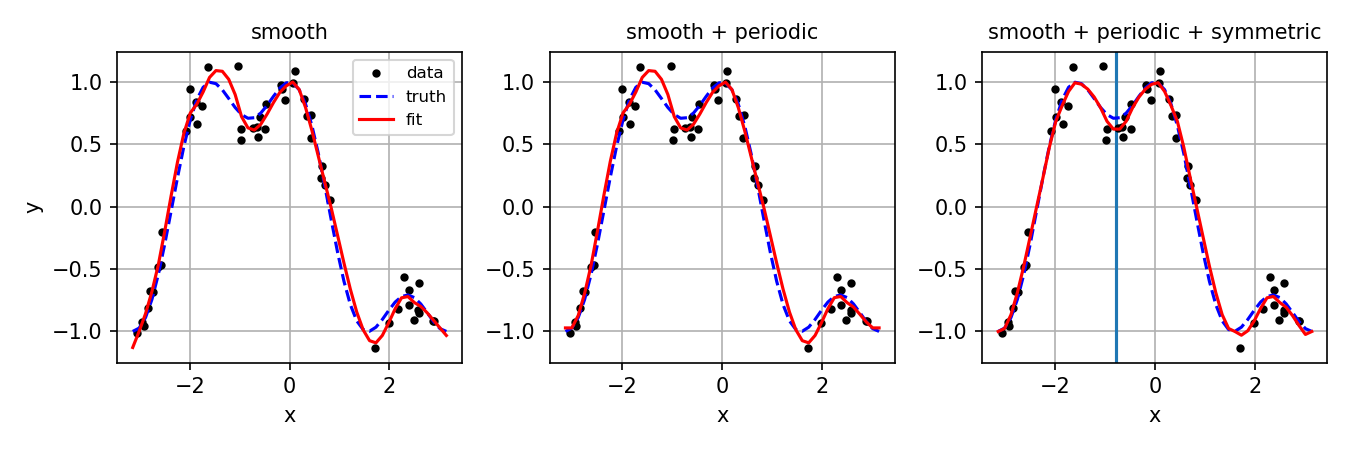}
    \caption{Fitting data generated around the true function $f(x)=\cos ^3(x)-\sin ^3(x)$ in the range $x \in [-\pi,\pi]$ with three different priors.}
    \label{fig:symmetric_demo}
\end{figure}

\section{Monotonic functions}

In some applications, one might want to require that the function is monotonically increasing or decreasing (along some input dimension). Such prior information cannot be directly included into our linear-Gaussian framework. One way to include monotonicity constraints is to re-write the estimation problem as a constrained linear-Gaussian problem. Efficient quadratic programming methods, such as \cite{goldfarb83}, exist for solving such problems.

Let us consider the constrained linear-Gaussian system
\begin{align}
\mathbf{y} &=  \mathbf{A}\pmb{\theta} + \pmb{\varepsilon} \\ 
\pmb{B\theta} &\sim \mathrm{N} \left( \pmb{\mu}_{\mathrm{pr}}, \pmb{\Gamma}_{\mathrm{pr}} \right) \\
\mathbf{C}\pmb{\theta} & \geq \mathbf{c},
\end{align}
where the last equation gives the constraints. As an example, monotonicity can be obtained by constraining the first order difference to be positive. Similarly, convexity could be obtained by requiring positivity from the second derivatives. Equality constraints can be included in the problem as well.

The above system is not quite in the format for which quadratic programming methods are directly applicable, but transforming the system to such a form is simple. The general quadratic programming problem format is 
\begin{align}
    \mathrm{minimize}\ \ \ & \frac{1}{2}\pmb{\theta}^T \mathbf{G} \pmb{\theta} - \mathbf{a}^T \pmb{\theta} \\
    \mathrm{so\ that}\ \ \ & \mathbf{K}^T \pmb{\theta} \geq \mathbf{b} .
\end{align}

It is easy to verify that our constrained linear Gaussian system can be transformed to the above format by setting $\mathbf{K} = \mathbf{C}^T$, $\mathbf{b} = \mathbf{c}$ and
\begin{align}
    \mathbf{G} &= \mathbf{A}^{\mathrm{T}} \pmb{\Gamma}_{\mathrm{obs}}^{-1} \mathbf{A} + \mathbf{B}^{\mathrm{T}} \pmb{\Gamma}_{\mathrm{pr}}^{-1} \mathbf{B} \\ 
    \mathbf{a} & = \mathbf{A}^{\mathrm{T}} \pmb{\Gamma}_{\mathrm{obs}}^{-1} \mathbf{y} + \mathbf{B}^{\mathrm{T}} \pmb{\Gamma}_{\mathrm{pr}}^{-1} \pmb{\mu}_{\mathrm{pr}}.
\end{align}

\begin{figure}
    \centering
    \includegraphics[width=0.6\textwidth]{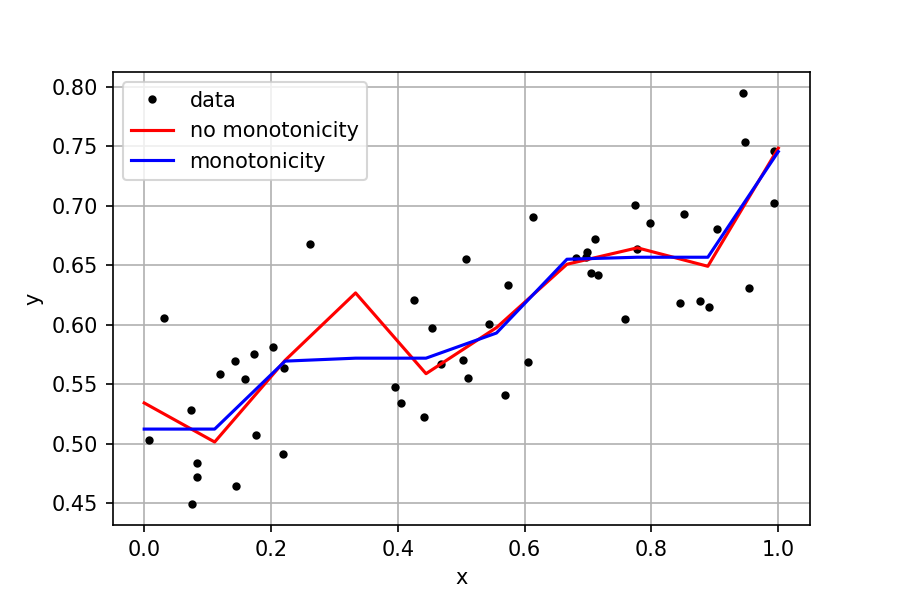}
    \caption{Fitting random data generated around $y=(1+x)/2$ with and without the monotonicity constraint.}
    \label{fig:monotonic_demo}
\end{figure}

A simple example of fitting monotonic functions is given in Fig. \ref{fig:monotonic_demo}. Note that this approach only gives the MAP estimate of the problem, not the whole posterior distribution nor the posterior mean estimate.

Another approach for including monotonicity information is described in \cite{riihimaki2010}, where monotonicity is obtained by placing synthetic derivative observations at chosen input locations. However, this approach is not directly applicable in the GAM context.

\section{Global vs. local bases: benefits and downsides}
\label{sec:benefits}

In this paper, we have presented two ways for formulating GAMs: the "global" Gaussian process approach and the "local" approach utilizing difference priors. One benefit of the GP approach is that one has a wide selection and literature of different kernel functions available. With the dimensionality reduction trick, if sufficient smoothness can be assumed, the method is also computationally efficient as input dimension and data amounts grow. In addition, it is explicitly clear what kind of functions the prior yields; sampling from the prior is straightforward, it is easy to control the range of allowed function values a priori, and there are no identifiability issues, since all functions are given their own mean values. Moreover, the GP method is discretization invariant by construction.

Another benefit of the GP approach with dimension reduction is that it can be plugged into nonlinear models without increasing the dimensionality of the estimation problem too much. For instance, Markov chain Monte Carlo (MCMC) sampling of the basis function weights can be much easier in the lower dimensional problem, as illustrated in \cite{marzouk, tukiainen16}.

Perodicity and symmetry are easy to include explicitly in the GP covariance functions, the local approach requires more manual work to set up the difference matrices appropriately. Then again, the difference priors offer more flexibility in setting the axis of symmetry, for instance.

The obvious downside of the GP approach is its global nature; it is hard to describe, e.g., sharp local features or to have spatially varying smoothess. Moreover, one needs to always take a stand about the level of the function values (mean), whereas in the local approach we can only penalize the derivatives and not impose any prior knowledge about the function values themselves. In addition, the extrapolation behavior of the GP method is always the same: the functions tend towards their priors outside the data range. This can be undesirable, and with the local approach one can obtain, e.g., linear or quadratic extrapolation behavior.

Finally, we wish to note that choosing the appropriate approach depends on the application at hand, and one cannot clearly state that one is better than the other.



\section{Hyperparameter estimation}
\label{sec:hyperpar}

In many problems, the modeller can be happy with manually found values for the hyperparameters (correlation lengths, difference prior variances, observation error variances, etc). In general, however, it's useful to have methods for automatic calibration of these values. This section discusses various approach for hyperparameter estimation in GAMs.

\subsection{Direct MAP estimation}

One straightforward way to get the hyperparameter values is to do a direct MAP optimization in the joint space of the parameters and the hyperparameters. This is obviously not the full solution to the Bayesian estimation problem, since it only gives us the posterior distribution of the parameters given the MAP estimate for the hyperparameters, but in practice this can lead to well working estimates. Note also that we can exploit the structure of the problem in the optimization; given the hyperparameters, the log-posterior is a quadratic function, for which the optimum can be found analytically. We can thus simply repeatedly fit the model within an iterative optimization loop.

\begin{figure}
    \centering
    \includegraphics[width=0.7\textwidth]{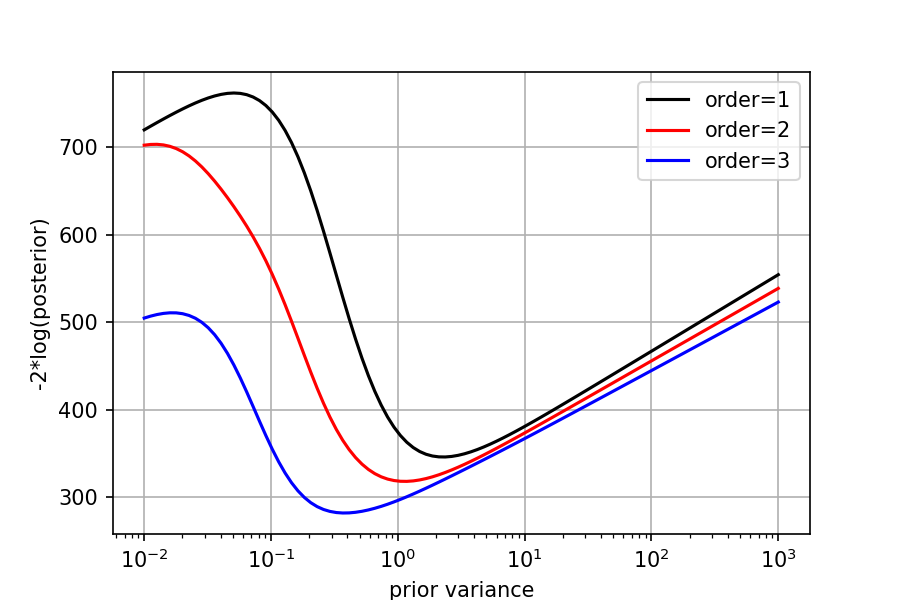}
    \caption{minimum negative log-posterior density values given prior variances for the problem illustrated in in Fig \ref{fig:diff_demo}.}
    \label{fig:hyperpar_demo}
\end{figure}

Let us consider, for instance, the simple fitting problem illustrated in Fig. \ref{fig:diff_demo}, where we used three orders for the difference priors and four different prior variances for each order. Here, fit the model repeatedly for a range of prior variances for each order and calculate the log-posterior value of the fitted model. The results are given in Fig. \ref{fig:hyperpar_demo}. One can see that there is a clear "optimal" prior variance value around 1 for each prior order.

Note that the posterior density has a pathological maximum when prior variance gets closer to zero. Zero prior variance makes the prior model (differences) fit to the data perfectly, which causes this corner case. The pathological optimum can be avoided by setting sensible limits for the prior variances (like in the example in Fig. \ref{fig:hyperpar_demo}), or to include other prior information about the variance parameters.

\subsection{Maximizing evidence}\label{sec:ev-max}

The idea in maximizing the evidence, a.k.a marginal likelihood, is to calculate the likelihood of the data under the prior distribution and maximize that. To derive the prior predictive distribution, let us recall our linear Gaussian system:
\begin{align*}
\mathbf{y} &=  \mathbf{A}\pmb{\theta} + \pmb{\varepsilon} \\ 
\mathbf{B}\pmb{\theta} &\sim \mathrm{N} \left( \pmb{\mu}_{\mathrm{pr}}, \pmb{\Gamma}_{\mathrm{pr}} \right).
\end{align*}
Assuming that the prior system in the lower equation is not under-determined, we can write the prior as $\pmb{\theta} \sim N(\pmb{\mu}_B, \pmb{\Sigma}_B)$, where
\begin{align}
\pmb{\mu}_B &= \pmb{\Sigma}_B \mathbf{B}^{\mathrm{T}} \pmb{\Gamma}_{\mathrm{pr}}^{-1} \pmb{\mu}_{pr} \\ 
\pmb{\Sigma}_B^{-1} &= \mathbf{B}^{\mathrm{T}} \pmb{\Gamma}_{\mathrm{pr}}^{-1} \mathbf{B} .
\end{align}

The prior predictive distribution is thus
\begin{equation}
    \mathbf{y} \sim \mathrm{N} \left( \mathbf{A} \left( \mathbf{B}^{\mathrm{T}} \pmb{\Gamma}_{\mathrm{pr}}^{-1} \mathbf{B} \right)^{-1} \mathbf{B}^{\mathrm{T}} \pmb{\Gamma}_{\mathrm{pr}}^{-1} \pmb{\mu}_{pr},\ \mathbf{A} \left( \mathbf{B}^{\mathrm{T}} \pmb{\Gamma}_{\mathrm{pr}}^{-1} \mathbf{B} \right)^{-1} \mathbf{A}^{\mathrm{T}} + \pmb{\Gamma}_{\mathrm{obs}} \right).
\end{equation}
Note that the above is computable only if the prior is proper, which might not be the case if the matrix $\pmb{B}$ is rank deficient, for instance. If we only penalize some derivative of the functions, we will end up in such a case. Then again, one can always add some non-informative priors for the function values and get this approach working.


\subsection{Cross-validation and posterior predictive score}

Cross-validation (CV) is another commonly used approach for hyperparameter tuning in Bayesian (and other) models. CV works by leaving out a part of the data, which is used to evaluate the goodness of the fit. The goodness-of-fit score here could be, e.g., a sum of squared deviations between the model predictions and the data. In the Bayesian setting, a preferable approach might be to use the "posterior predictive score", or, the likelihood of the left out data given the fitted model. In our linear-Gaussian setup, the posterior predictive distribution for new data $\mathbf{y*}$ is
\begin{equation}\label{eq:cv-post-pred}
    \mathbf{y*} | \mathbf{y}  \sim \mathrm{N} \left( \mathbf{A}\pmb{\mu}_{pos}, \mathbf{A}\pmb{\Gamma}_{\mathrm{pos}}\mathbf{A}^{\mathrm{T}} + \pmb{\Gamma}_{\mathrm{obs}} \right),
\end{equation}
and the cross-validation score is the log-density value of $\mathbf{y*}$ under this distribution.

Note that we often have only the (often sparse) inverse of the posterior covariance available from equation \ref{post}, so the above computation can be carried out via solving linear systems. While $\mathbf{A}$ and $\pmb{\Gamma}_{\mathrm{pos}}^{-1}$ are often sparse, the above covariance is dense, making the computation problematic if the number of observations $N$ is large because the size of the covariance in \eqref{eq:cv-post-pred} is $N \times N$.

Finally, the CV calculation works also if the prior is improper, unlike the evidence maximization approach discussed in \ref{sec:ev-max}.

A comparison of the MAP and posterior predictive scores for a simple fitting problem is given in Fig. \ref{fig:hyperpar_comp} below. Both give very similar results. Note that the evidence -based method was not applicable here due to a rank deficient prior system matrix $\mathbf{B}$.

\begin{figure}
    \centering
    \includegraphics[width=0.7\textwidth]{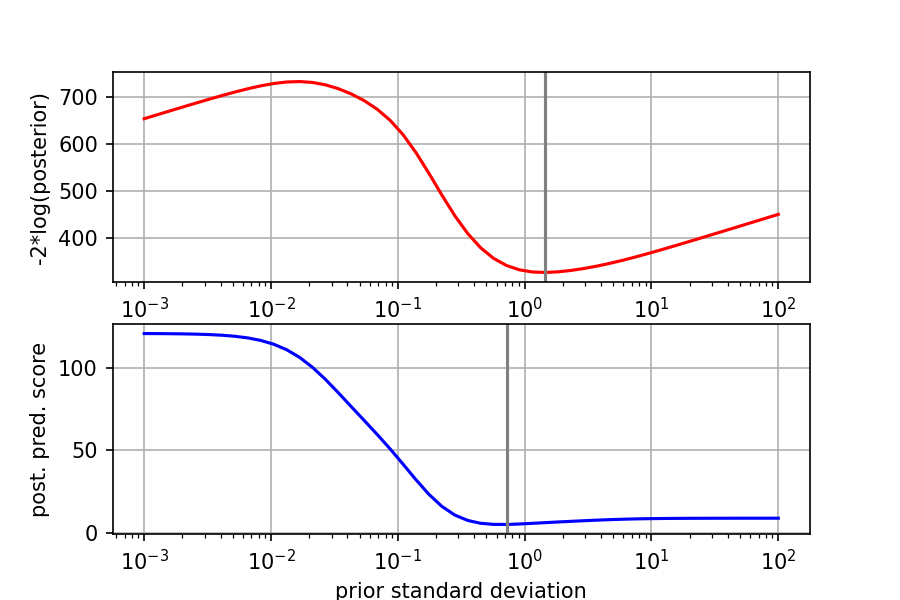}
    \caption{Comparison of two methods for hyperparameter calibration, using the simple fitting example illustrated in Fig. \ref{fig:diff_demo} using second order smoothness prior. Top: direct MAP estimation, bottom: posterior predictive score calculated for 10 data points left out from the model fitting.}
    \label{fig:hyperpar_comp}
\end{figure}

\subsection{Other Approaches}

As mentioned above, the log-posterior is often analytically available when the hyperparameters are fixed. The same can be true for the hyperparameters (e.g. in the local basis function case), given that suitable (e.g. conjugate) priors are used for them. This would enable possibly efficient block-wise optimization and Gibbs sampling schemes for optimizing the parameters and getting samples from the posterior.

Another possibility is to use variational inference techniques for estimating both the parameters and the hyperparameters. If conjugate priors are used, one could apply standard variational Bayesian updates to get an approximation of the posterior, see, e.g., \cite{bayespy} for discussion and a \texttt{Python} implementation of the approach. Also the INLA approach \cite{rue09} offers an efficient technique to approximate the posterior for both the parameters and the hyperparameters jointly.


\section{Further numerical examples}

The previous sections already contained various numerical illustration of the methodology presented here. In this section, we give a few more numerical examples. The code for these and the previous examples are made available \footnote{\url{https://github.com/solbes/gam_paper_examples}}. 

\subsection{Mauna Loa CO2 data}

Here, let us take a look at the famous Manua Loa CO2 dataset, which has monthly mean measurements of the CO2 level at the Mauna Loa observation station \footnote{Dr. Pieter Tans, NOAA/GML (\url{https://gml.noaa.gov/ccgg/trends/}) and Dr. Ralph Keeling, Scripps Institution of Oceanography (\url{https://scrippsco2.ucsd.edu}).}. The data shows a clear smooth trend in time, plus an annual periodic component. This motivates a model
\begin{equation*}
    y = f_1(\mathrm{time}) + f_2(\mathrm{month}) + \varepsilon,
\end{equation*}
where $f_1$ and $f_2$ are smooth functions, and the latter is periodic. 

\begin{figure}
    \centering
    \includegraphics[width=0.7\textwidth]{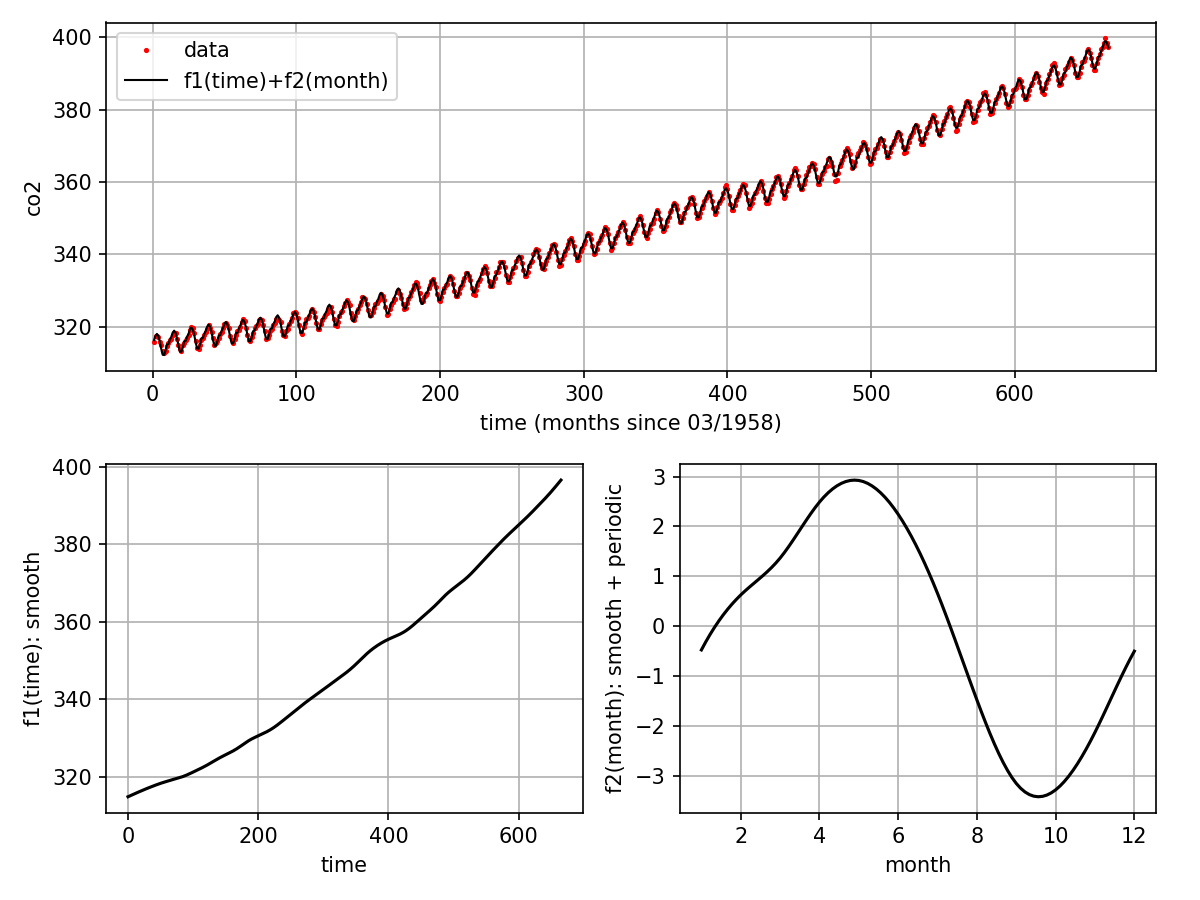}
    \caption{Top: data (red) vs. the fitted model (red). Bottom row: the fitted functions $f_1(\mathrm{time})$ and $f_2(\mathrm{month})$.}
    \label{fig:mauna_loa_manual}
\end{figure}

Let us first fit the data with manually chosen smoothness parameters, using the local basis function approach. See the attached codes for details about the chosen parameters. The results are given in Fig. \ref{fig:mauna_loa_manual}. One can observe that the model fits the data well and estimates good looking smooth trends.

\begin{figure}
    \centering
    \includegraphics[width=0.7\textwidth]{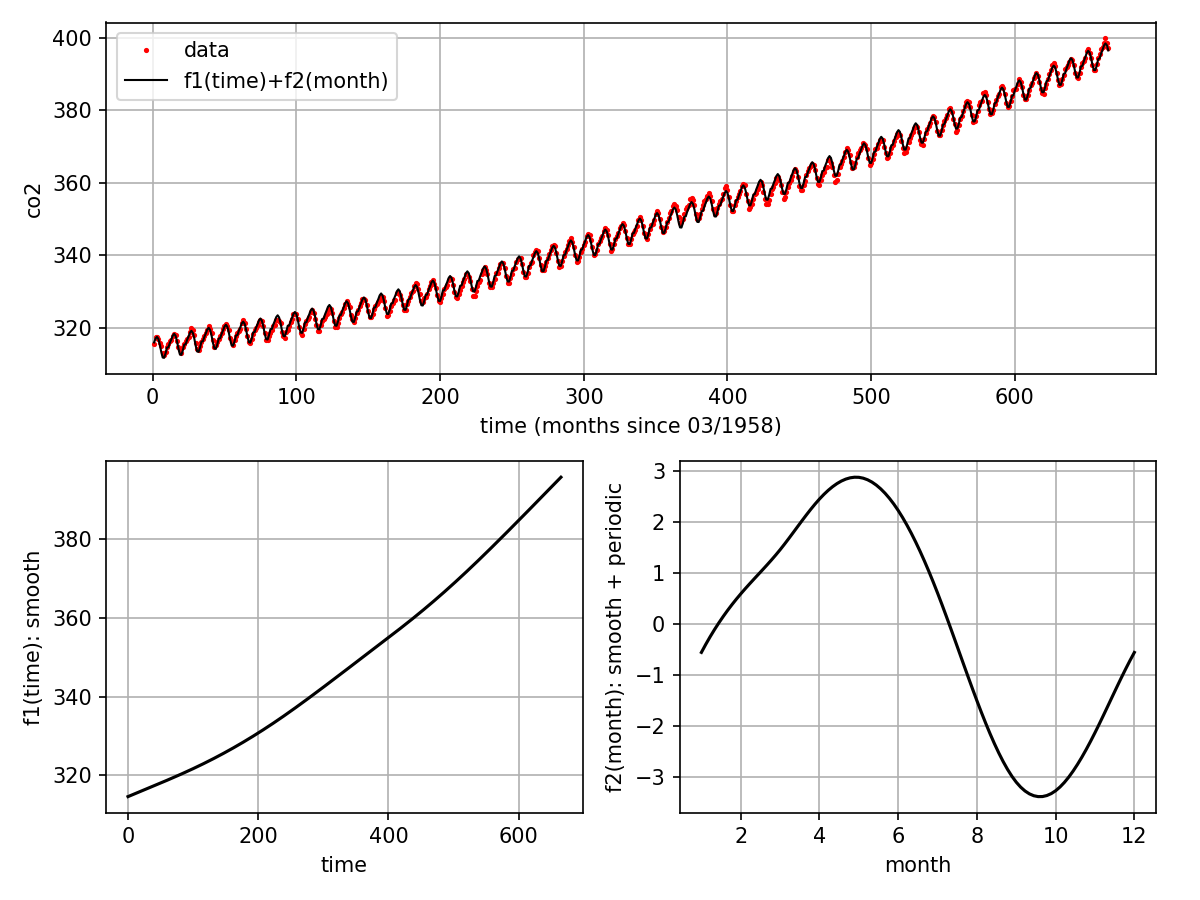}
    \caption{Top: data (red) vs. the fitted model (red). Bottom row: the fitted functions $f_1(\mathrm{time})$ and $f_2(\mathrm{month})$.}
    \label{fig:mauna_loa_auto}
\end{figure}

Let us then maximize the likelihood of the smoothness parameters for both smooth functions. The results are given in Fig. \ref{fig:mauna_loa_auto}. Compared to the previous fit, the function $f_1$ gets a bit smoother form, but otherwise the fit is similar.

\subsection{Synthetic example: hyperparameter tuning}

To demonstrate the various hyperparameter learning methods, let us consider a synthetic example here modified from \cite{wood17}. We generate random data in the interval $[x,y]$ by adding Gaussian noise to the true function
\begin{equation*}
    f_{true}(x) = 2\sin(\pi x) + \exp(2x) + x^{11}(10(1-x))^6/5+10^4x^3(1-x)^{10},
\end{equation*}
see the attached code for details of the experiment.

\begin{figure}
    \centering
    \includegraphics[width=0.7\textwidth]{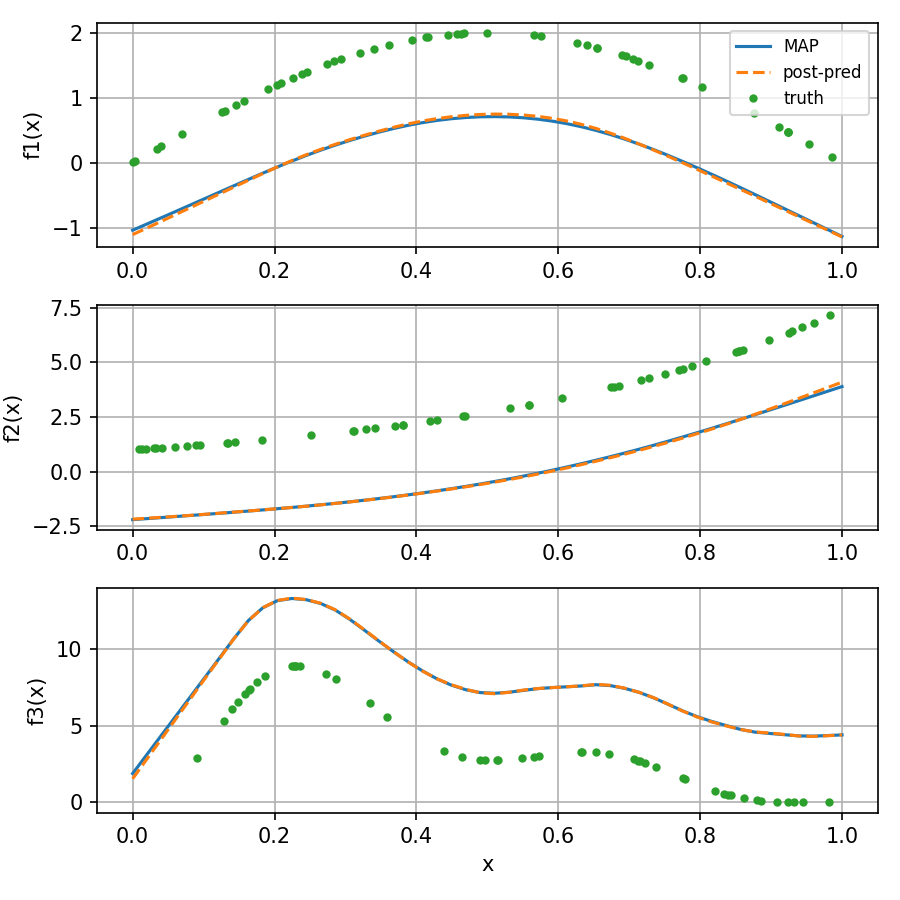}
    \caption{Simultaneous estimation of a the three component functions of a model using the local basis function approach. The smoothness level hyperparameters are estimated with two different tuning methods: combined MAP and posterior predictive maximization.}
    \label{fig:gam_demo_hyper}
\end{figure}

That is, we have three additive functions with different smoothness levels, and tuning the smoothness levels manually requires a lot of effort. Here, we fit a model
\begin{equation*}
    y = f_1(x)+f_2(x)+f_3(x)+\varepsilon
\end{equation*}
using the local basis function approach, and estimate the smoothness levels for all the three functions using the hyperparameter tuning methods presented in Section \ref{sec:hyperpar}. The data and the function fits for the individual components are given in Fig. \ref{fig:gam_demo_hyper}. Note that the shapes of the functions are recovered well, but the levels of the function values themselves are not identifiable from the data. Note also that the evidence -based method was not available in this example due to an improper prior.

\section{Conclusions}

Generalized additive models (GAMs) are flexible tools for fitting "semi-parametric" models, where parametric and non-parametric techniques are blended together. This paper discussed two different formulations for GAMs: the "global basis function approach", where the non-parametric functions in GAMs are described via Gaussian Processes (GPs), and a "local basis function approach", where the function values are directly treated as unknown parameters in the model, which are then penalized with Gaussian priors to achieve, e.g., suitable smoothness levels for the functions. Various numerical aspects of the model fitting were discussed, including dimensionality reduction, spatially varying smoothness, monotonic functions and hyperparameter tuning methods.

Which approach is the most suitable for practical modeling tasks depends on the problem at hand. The local basis function approach gives more flexibility in designing the shape of the functions, then again requiring more manual work in setting up the prior matrices correctly. The global GP approach can be computationally very efficient, when the dimension reduction trick can be applied, and can be more easily embedded into nonlinear models. See Section \ref{sec:benefits} for more discussion on the benefits and downsides of the two approaches.

As a final note, two \texttt{Python} packages for fitting such GAMs were developed along with this work. The first package, \texttt{gammy}\footnote{\url{https://github.com/malmgrek/gammy}}, is designed for fitting GAMs with the global (Gaussian process) basis function approach. The latter, \texttt{blinpy}\footnote{\url{https://github.com/solbes/blinpy}}, is a general tool for fitting Bayesian linear-Gaussian systems in the form described in Section \ref{sec:lingaus}, and contains various tools for constructing the system and prior matrices. Refer to the package pages for examples.




\bibliographystyle{unsrt}  
\bibliography{references}  

\end{document}